\documentclass[pre,twocolumn,preprintnumbers]{revtex4-1}
\usepackage{amsmath,bm,xspace,SIunits,braket,color,graphicx}

\begin{document}
\preprint{SAND 2013-9585J}
\title{First-principles calculation of entropy for liquid metals}
\date{\today}
\author{Michael P.~Desjarlais}
\email{mpdesja@sandia.gov}
\affiliation{Sandia National Laboratories, Albuquerque, New Mexico 87185}
\begin{abstract}
We demonstrate the accurate calculation of entropies and free energies for a variety of liquid metals using an extension of the two phase thermodynamic (2PT) model based on a decomposition of the velocity autocorrelation function into gas-like (hard sphere) and solid-like (harmonic) subsystems.  The hard sphere model for the gas-like component is shown to give systematically high entropies for liquid metals as a direct result of the unphysical Lorentzian high-frequency tail.  Using a memory function framework we derive a generally applicable velocity autocorrelation and frequency spectrum for the diffusive component which recovers the low frequency (long time) behavior of the hard sphere model while providing for realistic short time coherence and high frequency tails to the spectrum.  This approach provides a significant increase in the accuracy of the calculated entropies for liquid metals and is compared to ambient pressure data for liquid sodium, aluminum, gallium, tin, and iron.  The use of this method for the determination of melt boundaries is demonstrated with a calculation of the high pressure bcc melt boundary for sodium.
With the significantly improved accuracy available with the memory function treatment for softer interatomic potentials, the 2PT model for entropy calculations should find broader application in high energy density science, warm dense matter, planetary science, geophysics, and material science.
 \end{abstract}
\maketitle
\section{Introduction}
The first-principles calculation of entropies and free energies for liquids and the determination of liquid-solid phase boundaries is a long standing problem in molecular dynamics and one that has received considerable attention with a variety of approaches.  Whereas the total energy and pressure are readily available through direct calculation in classical or first-principles molecular dynamics codes, the calculation of the entropy remains the major impediment to construction of the Helmholtz or Gibbs free energies.  A rigorously justifiable and well established approach is that of thermodynamic integration~\cite{kirkwood,moriartyAlMelt, sugino, frenkelsmit}.  However, successful execution of this approach requires a considerable auxiliary architecture of tools, the choice of a good reference system for the problem at hand, and often numerous integration points along the thermodynamic integration path.  With respect to the determination of liquid-solid phase boundaries, the simulation of two phases in co-existence~\cite{morris94CoEx, alfe03CoEx,bonevCoExH, hernandezLi} and the monitoring of the boundary between them has produced a large body of results that agree well with experiment, albeit often requiring large system sizes and correspondingly long simulations for good convergence.  This method also does not provide direct access to the entropies or free energies of the two phases.  A direct melting method which has found widespread use in recent years is the Z method~\cite{belonoshko06Z,belonoshkoMgO,bouchet}. Carried out in the NVE ensemble, the Z method locates the melt boundary through superheating until melt and observing the corresponding pressure and temperature after relaxation to the liquid.  However, as the method bounds the melt curve from above, long simulations and corrections for finite simulation times might be needed to achieve the required precision~\cite{alfeZ}.  The vibration-transit liquid theory of Wallace and colleagues has provided considerable insight into the entropy change on melting through the characterization of the liquid as vibrations of the nuclei in random valleys interspersed with transits to neighboring valleys~\cite{wallace91c, wallace97b,wallacebook,wallace09,wallace10}. A relatively new and promising approach to the direct calculation of entropies and free energies for liquids is the two-phase thermodynamic (2PT) method of Lin {\it et al.}~\cite{lin2PT}.  This method uses only the primary simulation to compute the entropy and free energy through a decomposition of the vibrational density of states into solid-like and gas-like subsystems. 
The 2PT method was developed and validated within the Lennard-Jones system, and has subsequently been successfully applied to a wide variety of systems~\cite{2PTwater,2PTCO2, 2PTmixtures}.  In the context of liquid metals, the method has also been applied, but a significant overestimate of the entropy is apparent~\cite{TnB}.  In this work we develop a memory function representation of the gas-like subsystem that largely reduces the errors for liquid metals to the 1\% level while encompassing the earlier applications in a more general framework.  In Sec.~II we present an overview of the 2PT method.  The application of the 2PT method to liquid sodium is presented in Sec.~III and the source of the excess entropy illustrated.   A memory function representation for the gas-like component is developed in Sec.~IV and validated against experimental data for sodium, aluminum, tin, gallium, and iron in Sec.~V.   The memory function extension of the 2PT method is applied to calculating the high pressure bcc melt curve for sodium in Sec.~VI, where we find very good agreement with an alternative computational approach. 

\section{Overview of the 2PT method}
The principal quantity of interest in the 2PT method is the velocity autocorrelation (VAC) function and associated frequency spectrum derived from the velocity time histories  of $N$ atoms
\begin{equation}
\Phi(t)=\biggl\langle\frac{\sum_{i=1}^N{\bf v}_i(t)\cdot{\bf v}_i(0)}{\sum_{i=1}^N{\bf v}_i(0)\cdot{\bf v}_i(0)}\biggr\rangle_{T_{\text{w}}},
\label{vacfunc}
\end{equation}
\begin{equation}
F(\nu)=\int_0^\infty \Phi(t) \cos(2 \pi\nu t) dt.
\end{equation}
The averaging in Eq.~\eqref{vacfunc} is performed over several hundred uncorrelated time windows of length $T_{\text{w}}$.
Following Lin {\it et al.}, we decompose the total correlation function into solid-like and gas-like components.
\begin{equation}
\label{decompeqn}
\Phi(t)=(1-f_g)\Phi_s(t) + f_g \Phi_g(t),
\end{equation}
where $f_g$ is the gas-like fraction of the system, and the normalization is such that $\Phi_s(0) = \Phi_g(0) =1$.
Correspondingly
\begin{equation}
F(\nu)=(1-f_g)F_s(\nu) + f_g F_g(\nu).
\end{equation}
The zero frequency limit, $F(0)$, is directly related to the diffusion coefficient through $D = ({k}\,T/m) F(0)$, where $k$ is Boltzmann's constant, $T$ is the temperature, and $m$ is the mass.
Note that $F(\nu)$ as defined here is equivalent to $S(\nu)/12 N$ in Ref.~[\onlinecite{lin2PT}].  The unit of time for all computations is 10$^{-15}$ s and the corresponding frequency unit is 10$^{15}$ s$^{-1}$.
The normalization is such that $12 F(\nu)$ is the vibrational density of states and
$$\int_0^\infty 12 F(\nu) d\nu = 3,$$ satisfying the sum rule on the total density of states.  The same normalization applies independently to the solid-like and gas-like subcomponents
$F_s$ and $F_g$.

In order to perform the decomposition, Lin {\it et al.} modeled the gas-like portion as a hard sphere (HS) system~\cite{lin2PT}.
In the notation of Lin {\it et al.}, we define the hard sphere packing fraction of the gas-like component  $\gamma = f_g^{\text{HS}} y$, where $f_g^{\text{HS}}$ is the hard sphere fluidity factor, or gas-like fraction,
and $y$ is the hard sphere packing fraction for $N$ atoms in a simulation volume $\Omega$.
The fluidity factor $f_g^{\text{HS}}$ and the hard sphere packing fraction $y$ are related through $f_g^{\text{HS}} =  y^{2/3} \Delta$, where~\cite{lin2PT} 
\begin{equation}
\Delta =\frac{8}{3} F(0) \sqrt{\frac{\pi\,{k}T}{m}}
   \left(\frac{N}{\Omega }\right)^{1/3} \left(\frac{6}{\pi }\right)^{2/3}.
\end{equation}
With $\Delta$ determined by the system parameters and the zero frequency limit of the autocorrelation spectrum, 
and rewriting Eq.~(31) of Ref.~[\onlinecite{lin2PT}] in terms of $\gamma$,
we solve for $\gamma$
as the solution to
\begin{equation}
\frac{2 (1-\gamma )^3}{2-\gamma }-\gamma ^{2/5} \Delta ^{3/5}=0.
\end{equation}
The gas-like fraction is then
\begin{equation}
f_g^{\text{HS}}=\ \gamma ^{2/5} \Delta ^{3/5}.
\end{equation}
The resulting frequency spectrum for the gas-like component in the hard sphere model has the Lorentzian form
\begin{equation}
f_g^{\text{HS}} F_g^{\text{HS}}(\nu) = \frac{F(0)}{1+ (2\pi\nu/\alpha)^2},
\end{equation}
where $\alpha = f_g^{\text{HS}}/F(0)$.
The total VAC frequency spectrum is then decomposed into this gas-like component and the solid-like remainder.  The total ionic entropy $S_\text{ionic} = S_s + S_g$ 
may be written as integrals over the two components $(1-f_g^{\text{HS}})12 F_s(\nu) \equiv 12[F(\nu) - f_g^{\text{HS}}F_g^{\text{HS}}(\nu)] $ and $f_g^{\text{HS}} 12 F_g^{\text{HS}}(\nu)$, each with its appropriate weighting function~\cite{lin2PT}
\begin{align}
S_s&=N k \int_0^\infty  12[F(\nu) - f_g^{\text{HS}}F_g^{\text{HS}}(\nu)] W_s(\nu) d\nu, \\
S_g&=f_g^{\text{HS}} N k \int_0^\infty  12 F_g^{\text{HS}}(\nu) W^{\text{HS}}_g  d\nu,
\end{align}
where the weighting functions for the solid-like and gas-like components are given by
\begin{equation}
W_s(\nu)=\frac{h\nu/{k T}}{\exp(h\nu/{k T})-1} -\ln[1-\exp(-h\nu/{k T})]
\end{equation}
and
\begin{equation}
W^{\text{HS}}_g=\frac{1}{3}\left\{\frac{S^{\text{IG}}}{k}+ \ln\left[\frac{1+\gamma+\gamma^2 - \gamma^3}{(1-\gamma)^3}\right] + \frac{\gamma(3\gamma - 4)}{(1-\gamma)^2}\right\}
\end{equation}
respectively.  
The ideal gas component $S^{\text{IG}}$ is given by
\begin{equation}
\label{sidealgas}
\frac{S^{\text{IG}}}{k} = \frac{5}{2}+\ln\left[\left({\frac{2\pi m k T}{h^2}}\right)^{3/2}\frac{\Omega}{f_g^{\text{HS}}N}\right].
\end{equation} 
Given that the gas-like weighting function is independent of frequency, we can use the normalization of the
spectral components to write more simply
\begin{equation}
S_g= 3 N k f_g^{\text{HS}}W^{\text{HS}}_g.
\end{equation}

\section{Demonstration of the 2PT method with liquid sodium}
\begin{figure}[!h]
  \centering
  \includegraphics[width=0.47 \textwidth]{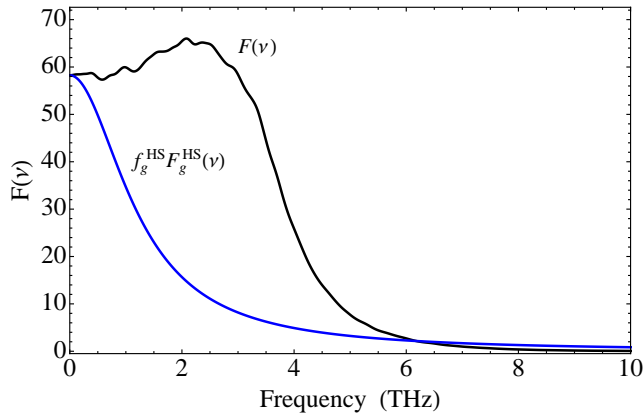}
  \caption{(Color online) The total VAC frequency spectrum for liquid sodium at 723\,K and ambient pressure is plotted (upper curve) along with the gas-like component from the hard sphere model (lower curve).}
  \label{fig:fnu_Na_723K}
\end{figure}

As an example application of this method to liquid metals, we choose liquid sodium at 723\,K along the ambient isobar.  This system has previously been treated by Teweldeberhan and Bonev~\cite{TnB}.  The calculations were performed using finite temperature density functional theory (FT-DFT)~\cite{mermin} with VASP~\cite{avasp, cvasp}.  The dynamics are performed in the Born-Oppenheimer approximation using the NVT ensemble with simple velocity scaling for the thermostat
\footnote{Test calculations with a Nos\'e-Hoover thermostat showed little difference, provided the thermostat frequency was appropriately chosen around the peak density of states.}.
For comparison with ambient pressure data, the volume is adjusted as the temperature is varied to provide average pressures within 5 kbar of zero. The resulting entropy computations were found to be insensitive to these deviations from the nominal ambient pressure volume.  The exchange-correlation functional chosen was generally the one that gave best agreement with known liquid densities at ambient pressure. The exchange-correlation chosen for the sodium calculations was AM05~\cite{am05} and the sodium atom is represented with a PAW potential~\cite{blochl, vasppaw}.  We used 128 atoms in the supercell and the Baldereschi~\cite{bald} mean value {\bf k}-point $\{\frac{1}{4}, \frac{1}{4}, \frac{1}{4}\} $ was used for sampling the Brillouin zone. The plane wave cutoff energy was 6 Ry.  The timestep was 2.0 fs and the simulations ran for 30\,000 timesteps.   The resulting autocorrelation frequency spectrum is shown in Fig.~\ref{fig:fnu_Na_723K} along with the gas-like component derived with a hard sphere model.  Note that for this liquid metal system the high frequency tail of the hard-sphere component exceeds the total frequency spectrum.

\begin{figure}[!h]
  \centering
  \includegraphics[width=0.47 \textwidth]{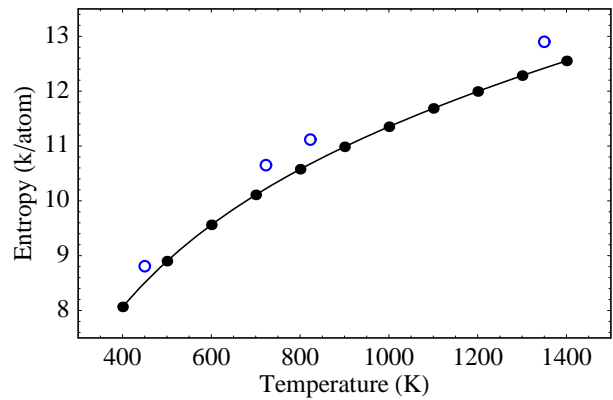}
  \caption{(Color online) Data and calculation results for liquid sodium.  Data from Hultgren {\it et al.}~are shown as filled circles and interpolated with the solid line.  The results of the 2PT method using a hard sphere model for the gas-like component are shown in open circles for 450\,K, 723\,K, 823\,K and 1350\,K.}
  \label{fig:NaDataHS}
\end{figure}

The entropies calculated for this system, using the 2PT method of Lin {\it et al.} outlined above, are shown in Fig.~\ref{fig:NaDataHS} for 450\,K, 723\,K, 823\,K and 1350\,K along with data from Hultgren {\it et al.}~\cite{hultgren}. 
Here, and in all total entropy comparisons with data in the sections that follow, the electronic contribution to the entropy $S_\text{e} = -k \langle\sum_i [f_i \ln f_i + (1-f_i)\ln (1-f_i)]\rangle$, where $f_i$ is the Fermi occupation of the $i$th electronic state computed in the course of the FT-DFT electronic minimization at each ionic timestep, is added to the ionic contribution to produce the total for comparison with data.  (Total entropy contributions that are not included in the reference data, such as isotopic entropy or nuclear spin entropy are not considered.)
 While the trend in the total entropy relative to the data in Fig.~\ref{fig:NaDataHS} is well characterized, the calculations yield systematically high entropies which exceed the data by approximately $0.4\,k/{\rm atom}$.  Given that the nominal change in entropy at constant volume for normal melting is around  $0.8\,k/{\rm atom}$~\cite{wallace91c}, this difference is significant.  Our results for the 2PT method are in good agreement with those of Teweldeberhan and Bonev~\cite{TnB} for the same system.

\begin{figure}[!h]
  \centering
  \includegraphics[width=0.47 \textwidth]{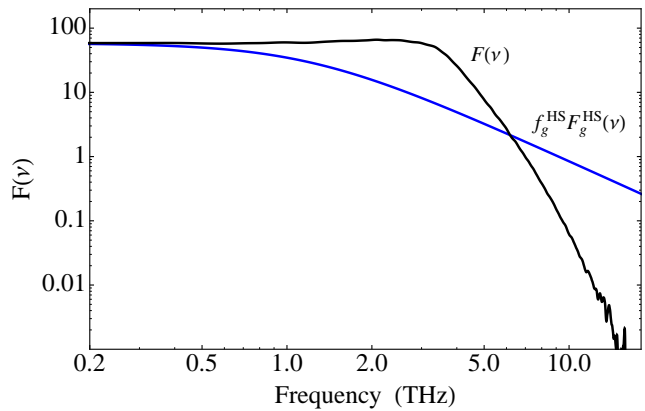}
  \caption{(Color online) The total VAC frequency spectrum for liquid sodium at 723\,K and ambient pressure is plotted on a log-log scale along with the gas-like component from the hard sphere model.}
  \label{fig:fnu_HS_log}
\end{figure}

The excess entropy may be traced to the long Lorentzian tail of the hard sphere system, which is more evident in the log plot shown in Fig.~\ref{fig:fnu_HS_log}.  Note that in this limit the weighting function for the gas-like subsystem is generally several times that of the solid-like component~\cite{lin2PT}.  This long tail is a direct result of the immediate exponential decay of velocity autocorrelation for hard spheres.  However, this exponential decay exaggerates the short time decay of correlations for the softer potentials found in liquid metals where a considerable coherence time is evident in the velocity autocorrelation time history.

\section{A generalized description of the gas-like component}
Our starting point for improving on the hard sphere frequency spectrum is a Memory Function (MF) representation~\cite{tjon,berne,singwi} of the velocity autocorrelation function
corresponding to the gas-like subsystem
\begin{equation}
\label{memfunc}
\frac{d\Phi_g(t)}{dt}=-\int_0^t K_g(\tau) \Phi_g(t-\tau) d\tau.
\end{equation}
The frequency spectrum can be calculated by first integrating Eq.\eqref{memfunc} for a given MF kernel $K_g(\tau)$ and performing the Fourier transform.
However, alternatively, and more expediently for our purposes, the frequency spectrum can be extracted directly from the MF kernel~\cite{singwi}
\begin{equation}
\label{fnu_laplace}
F_g(\nu)=\frac{1}{2}\bigg[\frac{1}{\hat{K_g}(i 2 \pi \nu)+ i 2\pi\nu}+\frac{1}{\hat{K_g}(-i 2 \pi \nu)- i 2\pi\nu}\bigg],
\end{equation}
where $\hat{K_g}(s)$ is the Laplace transform of $K_g(\tau)$.  

For the MF kernel we choose a Gaussian form, both for simplicity and because it
satisfies the required properties~\cite{berne} of being even in $\tau$ and having zero derivative at $\tau =0$.  The Gaussian form also guarantees that
higher order frequency moments of the resulting frequency spectrum exist~\cite{singwi}.
For a Gaussian Memory Function in the form
\begin{equation}
K_g(\tau)= A_g e^{-B_g \tau^2},
\end{equation}
we have
\begin{equation}
\label{k_laplace}
\hat{K_g}(s)= A_g\sqrt{\frac{\pi}{4 B_g}} \exp\bigg[\frac{s^2}{4B_g}\bigg]\ {\rm Erfc}\bigg[\frac{s}{2\sqrt B_g}\bigg],
\end{equation} \\
where ${\rm Erfc}[z]$ is the Complementary Error Function.  The zero frequency limit then provides a relation between A and B
\begin{equation}
\label{fgzero}
F_g(0)=\frac{1}{\hat{K_g}(0)}=\frac{1}{A_g}\sqrt{\frac{4B_g}{\pi}} = \frac{F(0)}{f_g}
\end{equation}
Requiring that the solution have the same low frequency (long time) behavior as the hard sphere system provides
a second relation between $A$ and $B$,
\begin{equation}
\label{abeqn}
A_g = 4 B_g/\left[2 + \sqrt{\pi(1 + 4 B_g/\alpha^2)}\right].
\end{equation}
Combining Eqs.~\eqref{abeqn}, \eqref{fgzero}, and \eqref{k_laplace}, and computing $F_g(\nu)$ with Eq.~\eqref{fnu_laplace}
provides a family of frequency spectra for the gas-like component, parametrized by $B_g$, the strength of the memory decay,
each member of which has the same zero frequency limit and low-frequency behavior as the hard sphere system.
A natural scale by which to measure the memory decay is the overall $1/e$ decay time $\tau_c$ defined by $\Phi(\tau_c)= 1/e$.
As $B_g \tau^2_c \to \infty$, the memory function becomes a delta function and the hard sphere result is recovered.  
For values of $B_g \tau^2_c /\pi^2 \lesssim 1$, the high frequency tail is considerably depressed
relative to the hard sphere limit.  In the following, a formal procedure for determining appropriate values of $A_g$, $B_g$, and $f_g$ from the velocity autocorrelation function is described.

In general, the autocorrelation may be written as a time series constructed from the even moments of the frequency spectrum~\cite{tjon,singwi}.
\begin{equation}
\label{serieseqn}
\Phi(t)= \sum_{n=0}^{\infty} (-1)^n \frac{M_{2n}}{(2n)!}t^{2n},
\end{equation}
where $M_{2n} = \langle\omega^{2n}\rangle $.  These moments are easily extracted from the total frequency spectrum, although higher order moments
are increasingly affected by noise in the high frequency tail.

In terms of the decomposition suggested in Eq.~\eqref{decompeqn}, we can collect the second and fourth moments as follows.
\begin{eqnarray}
\label{momentdecomp}
M_2 = (1 - f_g) M_{2s} + f_g M_{2g}&,\nonumber\\
M_4 = (1 - f_g) M_{4s} + f_g M_{4g}&.
\end{eqnarray}

The moments are related through the recurrence relation~\cite{tjon}
\begin{equation}
\label{recureqn}
M_{2n}= \sum_{j=1}^{n} (-1)^{j+1}  K_0^{(2j-2)} M_{2(n-j)},
\end{equation}
where
\begin{equation}
K_0^{(n)}= \frac{d^n}{dt^n} K(t)\bigg|_{t=0}.
\end{equation}
For Gaussian kernels of the form $A\exp({- B \tau^2)}$, these relations yield
\begin{eqnarray}
M_2& =&A,\nonumber \\
M_4& =&A^2  + 2 A B.
\end{eqnarray}

In the Memory Function representation, we can consider each component as evolving with respect to its own Gaussian Memory Function.
\begin{eqnarray}
\frac{d\Phi_s(t)}{dt}=-\int_0^t A_s e^{-B_s \tau^2} \Phi_s(t-\tau) d\tau&, \\
\frac{d\Phi_g(t)}{dt}=-\int_0^t A_g e^{-B_g \tau^2}\Phi_g(t-\tau) d\tau&,
\end{eqnarray}
\par
For the solid-like component, which has zero diffusion by definition, $F_s(0)=0$, so
$B_s =0$ follows from the analog of Eq.~\eqref{fgzero} for the solid-like component.
Inserting the respective moments for the solid-like and gas-like subsystems into Eqs.~\eqref{momentdecomp},
we obtain
\begin{align}
M_2 &=(1 - f_g) A_s + f_g A_g,\nonumber\\
M_4 &=(1 - f_g) A_s^2 + f_g (A_g^2 + 2 A_g B_g).
\end{align} 
\par

Using the first equation to eliminate $A_s$ in the second, provides a relation between $A_g$, $B_g$, and $f_g$ in terms of the
computed quantities $M_2$ and $M_4$.
\begin{equation}
\label{bgeqn}
B_g=\frac{{f_g} \left(2 M_2 {A_g}-M_4-{A_g}^2\right)+M_4-M_2^2}{2 {f_g} (1-{f_g}) A_g}
\end{equation}.

\begin{figure}[!h]
  \centering
  \includegraphics[width=0.47 \textwidth]{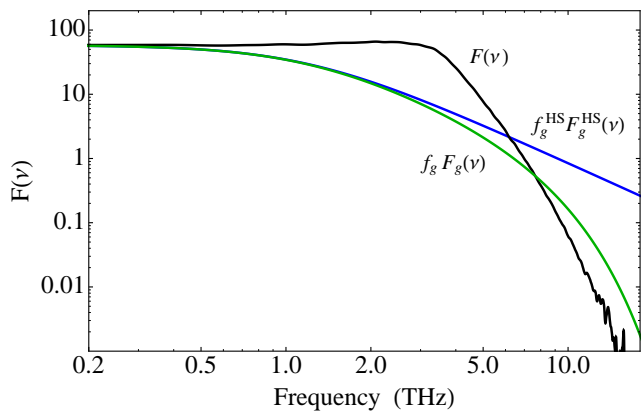}
  \caption{(Color online) The density of states for sodium at 723\,K.  The MF solution for the gas-like component was generated using the even moments $M_2\ {\rm and}\ M_4$.}
  \label{fig:fnu_E1_exmpl}
\end{figure}

We combine this with Eq.~\eqref{fgzero} to enforce the correct $F(0)$ limit, and with Eq.~\eqref{abeqn} to preserve the low frequency expansion found in the hard sphere system.  These three equations self-consistently determine the parameters of the Gaussian Memory Function for the gas-like component in terms of the first two even moments of the total frequency spectrum.  Note that at this level of description, the solid-like portion of the liquid is treated as an Einstein solid, in other words, characterized by a single frequency $A_s^{1/2}$.  The resulting MF solution is shown in Fig.~\ref{fig:fnu_E1_exmpl}.  Although the MF density of states for the gas-like component still exceeds the total distribution, there is already significant improvement over the hard sphere solution.  
Describing the solid-like portion with two frequencies requires 4 even moments.
\begin{align}
M_2 &=f_{s1} A_{s1}+ f_{s2} A_{s2} + f_g A_g,\nonumber\\
M_4 &=f_{s1} A^2_{s1} + f_{s2} A^2_{s2} + f_g (A_g^2 + 2 A_g B_g),\nonumber\\
M_6 &=f_{s1} A^3_{s1} + f_{s2} A^3_{s2} + f_g (A_g^3 + 4 A^2_g B_g + 12 A_g B^2_g),\nonumber\\
M_8 &=f_{s1} A^4_{s1} + f_{s2} A^4_{s2} \nonumber\\
&\ \ \ \ + f_g (A_g^4 + 6 A^3_g B_g + 28 A^2_g B^2_g + 120 A_g B^3_g ),
\end{align} 
with $ f_{s1} + f_{s2}  + f_g =1$.
\par
\begin{figure}[!h]
  \centering
  \includegraphics[width=0.47 \textwidth]{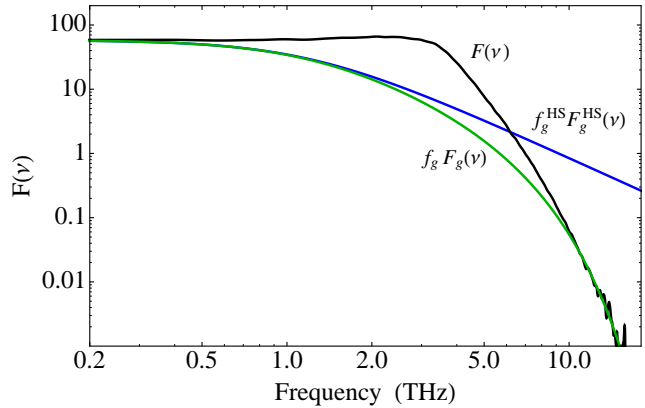}
  \caption{(Color online) The density of states for sodium at 723\,K.  The MF solution for the gas-like component was generated using four even moments $M_2, M_4, M_6, {\rm and}\ M_8$.}
  \label{fig:fnu_wt8_exmpl}
\end{figure}

The plot in Fig.~\ref{fig:fnu_wt8_exmpl} shows the MF solution with four even moments.  The self-consistent solution now accurately matches the tail of the total density of states.  This ability to smoothly match the high frequency limit further supports the choice of a Gaussian memory function.  An exponential memory function, in addition to violating the zero derivative property at $\tau =0$, would provide a high frequency tail with an $\omega^{-4}$ decay~\cite{berne}.  

It is important to note that the representation of the solid-like portion as one or more Einstein modes is an intermediate step in the solution for the gas-like component.  For the purpose of calculating the entropy, the true solid-like portion is still obtained by subtracting the gas-like component from the total and then applying the quasi-harmonic weighting function.

Having demonstrated the merging of the high frequency tail with the total distribution as a self-consistent solution at the level of four moments, we can, as a practical alternative to the formal approach of computing multiple moments and solving the systems of simultaneous equations, simply treat the proper high frequency behavior as a boundary condition and choose $B$ such that the high frequency tail of the gas-like component merges with the total density of states. This approach does require that the timestep is sufficiently small and the sampling extensive enough that the high frequency tail is not dominated by noise.  The determination of $B$ by matching to the high frequency tail is the approach taken in all the examples that follow and the entropy results so obtained are referred to as 2PT-MF.  

\begin{figure}[!h]
  \centering
  \includegraphics[width=0.47 \textwidth]{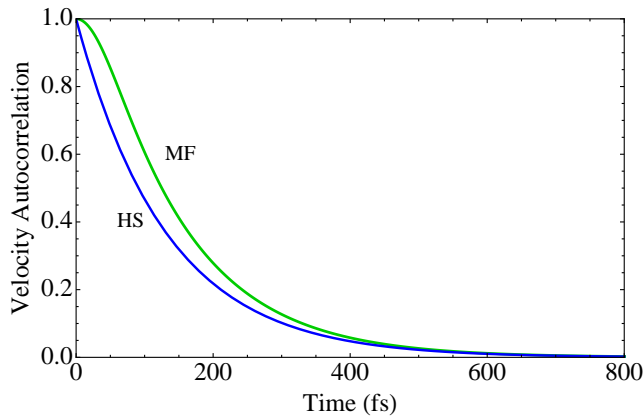}
  \caption{(Color online) The hard-sphere (lower curve) and the memory function (upper curve) autocorrelation functions in real time.  After a coherence time, the autocorrelation transitions to an exponential decay.}
  \label{fig:hs_mf_time}
\end{figure}

Figure~\ref{fig:hs_mf_time} shows the gas-like autocorrelation component in real time for the hard-sphere model (lower curve) and the real time solution to Eq.~(\ref{memfunc}) (upper curve).  
The hard-sphere autocorrelation is exponential, whereas the MF solution shows a realistic coherence for the early time evolution, giving over to an exponential decay later in time. 
This behavior can be roughly modeled as  $\exp[- \alpha t \exp(- \beta/t)]$ where $\beta$ is representative of a coherence time and $\alpha$ is the exponential decay found in the hard sphere model.
Having computed a self-consistent solution for the gas-like component with the proper high frequency behavior, the entropy is computed as in the 2PT treatment using the appropriate
weighting functions.  In the entropy weighting function for the gas-like component, the hard sphere correction to the ideal gas entropy is a function only of the total number of atoms,
the system volume, the temperature, and $F(0)$, or equivalently, the diffusion coefficient.  We use this correction to the ideal gas entropy weighting function without modification and calculate it in the hard-sphere limit.
However, the ideal gas contribution is modified to reflect the revised gas-like fraction $f_g$ in place of $f_g^{\text{HS}}$ in Eq.~(\ref{sidealgas}).

\section{Comparison of 2PT-MF calculations with data for several liquid metals}
\begin{figure}[!h]
  \centering
  \includegraphics[width=0.47 \textwidth]{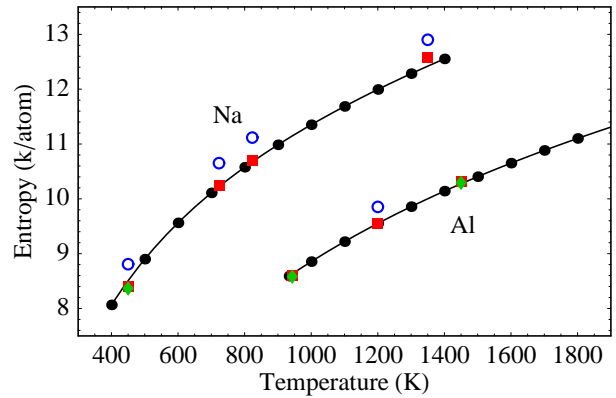}
  \caption{(Color online) Data and calculation results for liquid sodium and aluminum.  Data from Hultgren {\it et al.}~are shown in solid circles.  The results of the 2PT method using a hard sphere model for the gas-like component are shown in open circles. The results of the 2PT method with a Memory Function treatment of the gas-like component are shown in squares (128 Na atoms; 108 Al atoms) and diamonds (250 Na atoms; 256 Al atoms).}
  \label{fig:NaAlDataHSMF}
\end{figure}

Figure~\ref{fig:NaAlDataHSMF} shows the results of the MF treatment on the same sodium simulations as shown in Fig.~\ref{fig:NaDataHS} with the 2PT-MF solutions shown as squares.  The systematically better agreement with the data is clearly evident.  Note that the Hultgren data are the selected values taken from several data sets with varying degrees of uncertainty.  An additional simulation was performed with 250 atoms in the supercell for the 450\,K case (diamond), but no significant difference was found.  Also shown in Fig.~\ref{fig:NaAlDataHSMF} are data for aluminum (filled circles), a 2PT result (open circle), and 2PT-MF results (squares, 108 atoms; diamonds, 256 atoms).  For aluminum we used the PBE functional~\cite{pbe} and 2 fs timesteps.  The valence was $3s^2 3p^1$ and the plane wave cutoff was 17.7 Ry. The improvement with the MF method is substantial.  Size effects appear to be small.

\begin{figure}[!h]
  \centering
  \includegraphics[width=0.47 \textwidth]{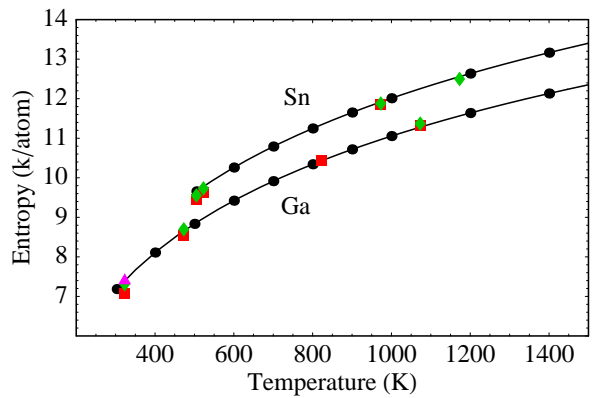}
  \caption{(Color online) Data and 2PT-MF calculation results for liquid tin (upper curve) and gallium (lower curve).  Data from Hultgren {\it et al.}~are shown in filled circles and interpolated with the solid line. Simulations were performed with 128 atoms (squares), 250 atoms (diamonds), and 432 atoms (triangle, Ga only).}
  \label{fig:GaSnDataMF}
\end{figure}

Figure~\ref{fig:GaSnDataMF} shows experimental data~\cite{hultgren} and 2PT-MF results for tin and gallium. 
Tin and gallium were chosen as potentially more challenging examples 
given their large negative correlation entropies, as defined and computed by Wallace~\cite{wallace91b,wallacebook}.
For tin and gallium we used the AM05 functional~\cite{am05} with a $4s^2 4p^2$ and $4s^2 4p^1$ valence respectively.  Simulations were performed 
with 128 atoms (squares), 250 atoms (diamonds), and 432 atoms (triangle, Ga only).  The total simulated times were 2.5 to 3 ps for 128 atoms, 2 ps for 250 atoms, and 1 ps for 432 atoms.  Here we do find significant size effects for temperatures close 
to melt where the correlation entropy is highest. The magnitude of the correlation entropies decay roughly linearly with $\ln(T/T_{\rm melt})$ with values 
at melt of approximately $-1.0\,k/{\rm atom}$ for tin and $-1.5\,k/{\rm atom}$ for gallium.  For tin at 973\,K and gallium at 1073\,K, 
the 2PT-MF results for 128 and 250 atoms nearly overlay.  Note that in case of both Sn and Ga, the 2PT-MF results approach the measured values from below with 
increasing cell size, suggesting that the smaller supercell exaggerates the negative correlation entropy for those materials that 
exhibit it.  Sodium and aluminum, in contrast to tin and gallium, have very little correlation entropy~\cite{wallace91b} and no significant size effects were found there.

\begin{figure}[!h]
  \centering
  \includegraphics[width=0.47 \textwidth]{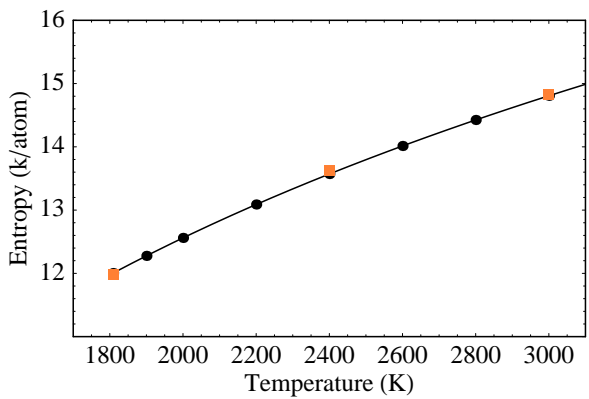}
  \caption{(Color online) Data from Hultgren {\it et al.}~for liquid iron are shown in filled circles.  Results of the 2PT-MF calculation results for liquid iron are shown with squares.  The iron simulations included the spin degree of freedom and were performed with 64 atoms in the supercell.}
  \label{fig:FeDataMF}
\end{figure}

The next material chosen for comparison is iron. Liquid iron at ambient pressure is expected to have a significant magnetic contribution to the entropy resulting from fluctuations in the local magnetization~\cite{grimvall89,grimvallbook}.  This adds an additional complication for iron and requires including the spin degree of freedom for a quantitative comparison with entropy measurements for liquid iron.  For the iron calculations we have used 64 atoms in the supercell with a $3d^6 4s^2$ valence and a plane wave cutoff energy of 29.4 Ry.  The total simulation times were 3~ps with timesteps of 0.5~fs. The exchange-correlation functional is PW91~\cite{pw91} and we used the Vosko-Wilk-Nusair~\cite{vwn}  interpolation for the correlation.  Following Grimval~\cite{grimvall89}, we estimate the magnetic contribution to the entropy from our spin polarized iron calculations by computing the RMS magnitude of the fluctuating amplitude of the local magnetic moments $\delta m_s$, in Bohr magnetons, averaged over time and the ion ensemble.  The magnetic entropy is then computed as
\begin{equation}
\label{magentropy}
S_{\rm mag}= k \ln\left[ 1 + \langle \delta m_s^2\rangle ^{1/2}\right],
\end{equation}
assuming completely disordered spin orientations in the liquid state.
The resulting values are on the order of $0.95$ to $1.02\,k/{\rm atom}$, in good agreement with estimates from data of approximately $1 \,k/{\rm atom}$ for iron~\cite{grimvall89,grimvallbook}, and are added to the electronic and ionic contributions to the entropy.  
The comparison of our 2PT-MF results with data for the total entropy of iron is shown in Fig.~\ref{fig:FeDataMF}.

\begin{figure}[!h]
  \centering
  \includegraphics[width=0.47 \textwidth]{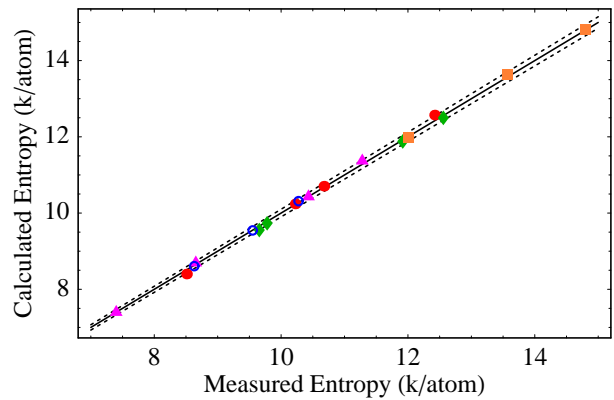}
  \caption{(Color online) A compilation of the entropy calculations for Na (filled circles), Al (open circles), Sn (diamonds), Ga (triangles), and Fe (squares), plotted against the measured values.  The dotted lines represent differences of 1\% above and below the measured values.}
  \label{fig:entropycomp}
\end{figure}

Figure~\ref{fig:entropycomp} shows a compilation of our 2PT-MF calculations plotted versus the selected values from Ref.~[\onlinecite{hultgren}].  The dotted lines indicate differences of 1\% above and below the selected measured values. The agreement is generally very good and overall shows no systematic differences with the data from experiments.

\section{Application to the BCC melt curve of sodium}

The complex high pressure melt curves and phase diagrams of the alkali metals have received much attention both theoretically and experimentally, and we have chosen the bcc melt curve of sodium as an example application of the 2PT-MF method for determining melt boundaries.  In addition to the experimental measurements of Zha and Boehler~\cite{zha} and Gregoryanz {\it et al.}~\cite{gregoryanz}, there have been numerous computational determinations of increasing degrees of sophistication~\cite{raty, hernandez, yamane, eshet}.  The two experimental investigations give widely different results that begin to diverge above 6 GPa.  The recent work of Eshet {\it et al.}~\cite{eshetNN, eshet} used a neural network (NN) algorithm and DFT calculations with the PBE functional and potentials with a $2s^2 2p^6 3s^1$ valence to generate accurate model potentials for sodium.  This enabled highly efficient classical MD simulations with many more atoms and longer simulation times than is feasible within the DFT molecular dynamics framework. Thermodynamic integrations from the Einstein crystal and Lennard-Jones liquid were used to determine the phase boundaries. The resulting phase diagram and melt boundaries provide an excellent computational benchmark for the 2PT method with the MF extension developed here for the same underlying functional.

Following the work of Hern{\' a}ndez and \'{I}\~{n}iguez~\cite{hernandez}, and Yamane {\it et al.}~\cite{yamane}, we have used a PAW potential with a $2p^6 3s^1$ valence for all calculations at pressures above 20 GPa to avoid unphysical dimerization in the liquid.  The lower pressure simulations included only the $3s^1$ electron in the valence. In keeping with the earlier work, and for direct comparison to the results of Ref.~[\onlinecite{eshet}], we have used the PBE functional for these calculations.
All calculations were performed with 128 atoms in the supercell and a $2\times2\times2$ Monkhorst-Pack grid~\cite{monkpack} of $\bf k$-points was used to sample the Brillouin zone.  The plane wave cutoff energies were 7.5 Ry and 24.3 Ry for the $3s^1$ and $2p^6 3s^1$ potentials respectively.  Simple velocity scaling at every timestep was used as a thermostat. 
For pressures below 20 GPa, simulations of 40 ps duration were performed with 1 fs timesteps.  For pressures above 20 GPa, simulations of 30 ps duration were performed with 0.5 fs timesteps.
These simulations in the NVT ensemble were carried out for the bcc and liquid phases at identical volumes and temperatures to obtain accurate total energies and in the case of the liquid to also compute the entropy as described here. For the purposes of demonstrating the accuracies of the liquid metal entropy calculation over a broad pressure range and avoid the potential for cancellation of errors between the liquid and solid computations, the entropies of the bcc phase were computed along isochors using a combination of quasi-harmonic phonon calculations with PHON~\cite{phonalfe} --- at sufficiently low temperatures  (100\,K to 200\,K) to avoid significant anharmonic contributions to the entropy --- followed by a thermodynamic integration in temperature along the isochor.  The anharmonic contributions to the bcc entropy at melt for the pressures examined here are found to be on the order of $(0.08  \pm 0.02)k$/atom, consistent with previous estimates for ambient conditions~\cite{wallace92}. The vibrational density of states for the solid and liquid were used to provide quantum corrections to the total energies and to the bcc entropy from thermodynamic integration. The resulting quantum corrections to the entropy are small, approximately $0.03 k$/atom, and largely cancel between the solid and liquid. Finally, the Helmholtz free energy was computed for each phase at various points along the isochor.  We locate the constant volume phase boundary by the crossing of the Helmholtz free energies as a function of temperature at $T_{melt}$ and, as the pressures and compressibilities between the two phases are generally close, we expand the Gibbs free energy locally to second order in $\delta V$ and compute the melt pressure for the corresponding Gibbs free energy crossing with the following relation.

\begin{equation}
P_{melt} = \frac{P_L \sqrt{\kappa_L} + P_S \sqrt{\kappa_S}}{\sqrt{\kappa_L} + \sqrt{\kappa_S}},
\end{equation}
where $P_{L,S}$ are the pressures and $\kappa_{L,S}$ are the isothermal compressibilities of the liquid and solid phases at $T_{melt}$ and the given volume.  For small volume changes and compressibilities that are comparable, this can be well approximated by $P_{melt} = (P_S + P_L)/2$~\cite{moriartyAlMelt}.

\begin{figure}[!h]
  \centering
  \includegraphics[width=0.47 \textwidth]{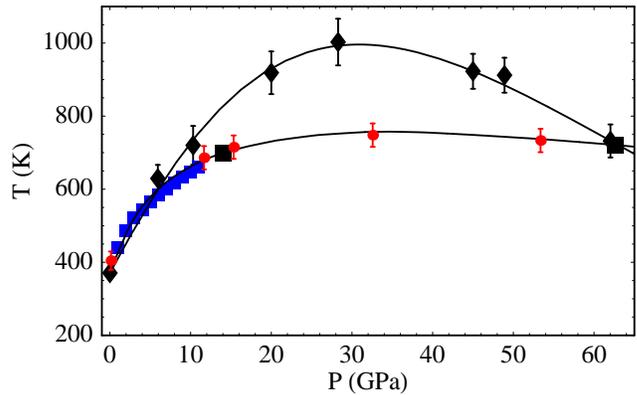}
  \caption{(Color online) The bcc melt curve of sodium. The diamonds are the experimental results from Ref.~[\onlinecite{gregoryanz}], the small squares at lower pressures are the earlier experimental results from Ref.~[\onlinecite{zha}].  The lower solid curve along with the two larger squares is the PBE/NN melt curve of Ref.~[\onlinecite{eshet}].  Results obtained with the 2PT-MF method are shown as solid circles.}
  \label{fig:namelt}
\end{figure}

The results of these calculations are shown in Fig.~\ref{fig:namelt}.  The diamonds are the experimental results of Gregoryanz {\it et al.}~\cite{gregoryanz}, the small squares at lower pressures are the earlier experimental results of Zha and Boehler~\cite{zha}.  The lower solid curve along with the two larger squares is the PBE/NN bcc melt curve of Eshet {\it et al}.~\cite{eshet}.  Results obtained with the 2PT-MF method are shown as solid circles.  The error bars shown for the 2PT-MF results reflect an estimated statistical uncertainty of approximately 0.4\% in the determination of the liquid entropy that results primarily from the finite run times and noise in the high frequency tail resulting from finite time steps.  No potentially underlying systematic errors, such as might result from the choice of exchange-correlation functional for example, are accounted for in the error bars.  Our results for the bcc melt curve are in very good agreement with the PBE/NN results of Ref.~[\onlinecite{eshet}], and the results of both approaches agree, perhaps fortuitously, with the earlier experimental results of Zha and Boehler and do not agree with the more recent results of Gregoryanz {\it et al.}

Having computed the entropies for the solid and liquid phases, we can make a direct connection to the vibration-transit (V-T) theory of melting~\cite{wallace91c,wallace97b,wallacebook,wallace09,wallace10}.
The V-T theory of melting has established that for monatomic liquids the per-atom constant volume entropy change at melt, defined as $\Delta S^*$, is a near universal constant for normal melting with a value of $(0.80 \pm 0.10)k$.  Furthermore, this near universal behavior reflects an underlying transit entropy contribution to the liquid entropy $S_{tr}(T/\theta_{tr})$, where $\theta_{tr}$ is a characteristic temperature. $S_{tr}(T/\theta_{tr})$ is a concave function and has a maximum value of around $0.8k$ for sodium, based on analysis of ambient pressure data.  For the melting of sodium at ambient pressure $\Delta S^* =0.73k~$\cite{wallace91c}.  For the 2PT-MF results shown in Fig.~\ref{fig:namelt}, the values of $\Delta S^*/k$, in order of increasing pressure, are 0.71, 0.79, 0.80, 0.77, and 0.78.  Aside from the value for the lowest pressure which is close to the known ambient value, the values for $\Delta S^*/k$ quickly increase to values near the known maximum value for sodium.  

In this context the maximum in the melt curve near 30~GPa and 1000\,K found in Ref.~[\onlinecite{gregoryanz}] is difficult to explain at the level of DFT with the PBE functional.  This would require a $\Delta S^*$ with an anomalously low value of around 0.5$k$.  Alternatively, if we assume a value of $\Delta S^*$ of  0.8$k$ at the apex, an increase of the liquid-bcc energy difference, relative to the DFT/PBE values, by around 38\% would be required to give agreement with experiment.

\section{Discussion}
An extension of the 2PT method~\cite{lin2PT} has been developed and implemented with particular emphasis on the application to liquid metals.
A memory function model for the diffusive gas-like component provides a natural framework for modeling the softer interactions and longer coherence times inherent in these systems.  Self-consistent solution of the model system provides a high frequency tail to the gas-like density of states that naturally merges with the total density of states obtained from long molecular dynamics simulations.  The method is tested on several liquid metals (Na, Al, Sn, Ga, Fe) and the agreement with the selected data values of Hultgren {\it et al.} is very good, generally agreeing within 1\%.  This extension of the 2PT method has been applied to the determination of the bcc melt boundary of sodium and very good agreement is found with highly converged simulations employing an alternative method~\cite{eshet}.  Without the memory function treatment of the diffusive component, the estimated peak of the melt curve from our calculations would be several hundred Kelvin lower as a consequence of the enhanced liquid entropy, and well below the point where the liquid in our simulations spontaneously freezes.

As the memory function extension to the 2PT method is completely general, it naturally encompasses the hard sphere limit and can be applied to any implementation of the 2PT method.  We have chosen to perform the entropy calculations here within a first-principles framework, motivated in large part by the direct access to the electronic and magnetic contributions to the entropy of liquid metals for quantitative comparison to the Hultgren data at the 1\% level.  However,  the memory function extension to the 2PT method is broadly applicable to molecular dynamics simulations of fluids with suitable model potentials. A distinct advantage of the 2PT or 2PT-MF method is the direct extraction of the entropy from the primary molecular dynamics simulation and straightforward post-processing of the velocity time history, albeit requiring longer simulation times for convergence than would be required for determining the pressure or total energy to the same level of accuracy.  As only one phase is simulated at a time, the approach should provide favorable size scaling relative to the simulation of two phases in coexistence.  The method should prove useful as an efficient means for locating release adiabats from high pressure liquid states achieved through shock compression, particularly when the release state of interest is far removed from the Hugoniot state. Recent work on planetary impacts considers just this scenario for silica~\cite{KrausSiO2}.  This direct approach to computing the entropy of liquid states, and the improvements developed here, should also prove useful in high energy density science, warm dense matter, planetary science, geophysics, physical chemistry, and material science.  While we have only demonstrated the method with monatomic metals, the 2PT approach is readily extended to molecular systems~\cite{2PTwater,2PTCO2} and mixtures~\cite{2PTmixtures,TnB, boatesMgO} and the memory function extension described here is equally applicable.

\bigskip
\section*{Acknowledgments}  The author gratefully thanks Luke Shulenburger, Martin French, and Thomas Mattsson for useful discussions.  All simulations were performed on Sandia National Laboratories High Performance Computing Platforms.  The sodium melt computations benefited substantially in efficiency from Paul Kent's {\bf k}-point parallelization scheme for VASP.  Sandia National Laboratories is a multi-program laboratory managed and operated by Sandia Corporation, a wholly owned subsidiary of Lockheed Martin Corporation, for the U.S. Department of Energy's National Nuclear Security Administration under contract DE-AC04-94AL85000.

\bibliographystyle{apsrev4-1}
\bibliography{entropybib}

\end{document}